# Effects of biaxial strain and local constant potential on electronic structure of monolayer SnSe

Feng Sun[1], Ting Luo[1], Lin Li[2], Aijun Hong[1*], Cailei Yuan[1*], Wei Zhang[3]

[1]*Jiangxi Key Laboratory of Nanomaterials and Sensors, School of Physics, Communication and Electronics, Jiangxi Normal University, Nanchang 330022, China*

[2] *Material Technology Institute, Yibin University, Yibin 644000, China*

[3] *State Key Laboratory of Hydroscience and Engineering, Department of Energy and Power Engineering, Tsinghua University, Beijing 100084, China*

Correspondence and requests for materials should be addressed to A.J.H. (email: 6312886haj@163.com or haj@jxnu.edu.cn) and C.L.Y. (email: clyuan@jxnu.edu.cn).

[**Abstract**]

We use the modified Becke-Johnson exchange potential (mBJ) with the spin-orbit coupling effect (SOC) to study effects of biaxial strain and local constant potential on electronic structure of monolayer SnSe. Our results show the fundamental band gap size can be tuned via biaxial strain. Compressive strain (tensile strain) can narrow (enlarge) band gap, and compressive strain causes the transition from quasi-direct to indirect band gap. Moreover, considering that any tuning of electronic structure is realized by changing the periodic potential distribution in the crystalline, we directly add constant potential (CP) to muffin-tin spheres. The results demonstrate that positive and negative CPs can narrow and enlarge band gap, respectively. At CP of 0.9 Ry, semiconductor-metal transition appears, and interestingly a new type of nearly linear dispersions occur at band edge. Our work is good for inspiring more experimental and further theoretical research works.

## I. INTRODUCTION

Since the discovery of graphene [1] a lot of focus has been put on exploring two-dimensional (2D) materials. A large amount of theoretical and experimental work shows 2D materials such as transition metal dichalcogenides (TMDs) [2], silicene [3], stanine [4] and black phosphorene (BP) [5] have great potential and promise in electronics and photonics applications. For instance, due to anisotropic physical properties and band gap tunability, BP as layered allotrope of phosphorus has received worldwide attention [6-8]. The IV-VI compounds (IV=Ge, Sn; VI=S, Se) [9,10] are semiconductors and possess layered puckered orthorhombic structure that is the same with BP (see Fig. 1). Some of their bulks have exhibited excellent performance in photovoltaic and thermoelectric applications [11-14]. Particularly, the single crystal SnSe in has been reported to be state-of-the-art thermoelectric material due to record high thermoelectric figure-of-merit (ZT = 2.6) at 923 K [15]. Recently, SnSe monolayer with lateral dimension of a few millimeters was synthesized via a two-step synthesis method [11], which offers a broad stage for theoretical and experimental researches.

Turning band gap and modifying band edge shape of 2D materials are interesting research topic and are the most concerned of band engineering. To this end, several strategies such as doping, strain, and electric field [16-23] have been used to tune electronic structures of 2D materials graphene, TMDs and so on. For example, by applying vertical electric field, the band gap of bilayer graphene opens is continuously tuned up to about 250 meV [24]. The electronic and magnetic properties of the BN zigzag nanoribbons are regulated by different percentages of hydrogenation [25], which renders BN-based nanomaterials possessing huge potential applications in novel integrated and functional nano-devices. Strain not only causes semiconductor-metal transition in transition metal dichalcogenide monolayers but also affect their magnetic properties [26,27]. Thus, it is interesting and necessary to study strain effect on electronic structure of the IV-VI compound SnSe monolayer. Moreover, as known, any band engineering strategies can realize to tune the electronic structure, which is based on that they modify the periodic potential distribution in the crystalline. Aside from the band engineering strategies, directly modifying the periodic potential distribution for obtaining willing electronic structure is worth studying.

In this work, for changing the periodic potential distribution, we first apply biaxial strain to monolayer cleaved from low-temperature phase of bulk SnSe. We find compressive strain (tensile strain) can narrow (enlarge) band gap. Subsequently, we directly modify the periodic potential distribution via applying constant potential (CP) to muffin-tin spheres. It is demonstrated that semiconductor-metal transition appears and a new type of nearly linear dispersions occur at band edge when the CP is at 0.9 Ry.

## II. METHOY AND DETAILS

In this work, we use full-potential density functional theory (FDFT) package WIEN2k [28] to perform optimization of the internal positions by minimization of the forces (1 mRy/a.u.) acting on the atoms. The calculations on electronic band structure are performed by using the Perdew-Burke-Ernzerhof (PBE) functional [29] with the modification of Becke-Johnson (mBJ) [30] as exchange-correlation potential, and the spin-orbit coupling effect (SOC) is taken into account. The cutoff parameter, total energy and charge convergences are respectively set to 7.0, $10^{-4}$ Ry and $10^{-4}$ e. The muffin-tin sphere radii of both Sn and Se atoms are fixed by 2.0 a.u.. In order to avoid layer-to-layer interaction, a 20 Å thick vacuum slab is built into the 2D structure. For structure optimization and electronic structure calculations, the $k$-mesh is set to $11\times11\times1$ and $51\times51\times1$, respectively. However, it changes to 1000 $k$-points along the special points for band structure calculations.

## III. RESULTS AND DISCUSSIONS
### A. biaxial strain effect

We first focus on whether the structure without strain has the minimum energy. Obviously, the total energy increases with increasing compressive or tensile ratio in Fig. 3(a), which is also evidence of our successful structural optimization. The PBE total energies are larger than the PBE+SOC. The total energy difference decreases with increasing strain ratio. It is worth mentioning that the mBJ total energy makes no sense due to no corresponding exchange and correlation functional for mBJ potential. Figs. 2 (b) and (c) show compressive strain (tensile strain) can narrow (enlarge) band gap, and the SOC effect can leads to small decrease of band gap in monolayer SnSe. The mBJ band gaps are larger than the PBE, which is acceptable because the normal PBE functional always underestimate the band gap and the mBJ scheme has high accuracy like GW method.

In order to studying biaxial strain effect on electronic structure, the mBJ band structure with respect to biaxial strain is calculated (see Fig. 3). For no strain case, we find that the monolayer SnSe belongs to a quasi-direct band gap semiconductor because the valance band maximum (VBM) and the conductivity band minimum (CBM) locate at almost the same position. There are two energetically almost degenerate VBMs respectively locating on the $\Gamma$-X and $\Gamma$-Y directions, which is the same case as the conduction band. This could be attributed to a certain degree of electron-hole symmetry especially along the $\Gamma$-X and $\Gamma$-Y directions. However, the symmetry is destroyed by the strain. For compressive strain, the decrease of band gap is due to high dispersion of valley bands, although the energy range of conductivity band (CB) becomes narrow with the increase of compressive strain ratio. While compressive strain ratio raise upon to 6%, the transition from quasi-direct to indirect band gap appears. The tensile strain induces the localization of valley band (VB) and thus leads to the increase of band gap. However, the tensile strain has slight effect on CB. By comparing Fig. 3 and Fig. 4, it is found that the splitting induced by the SOC is asymmetrical. The splitting of VB/CB along the $\Gamma$-Y direction is more obvious than along the $\Gamma$-X direction.

### B. constant potential effect

In the framework of the full potential linearized augmented plane wave (FP-LAPW) method [28,29,31], the unit cell is divided into (I) muffin-tin spheres and (II) interstitial region (see Fig. 5). We add the constant potential $V_{con}$ to Hamiltonian

in Kohn-Sham equations of density functional theory (DFT):

$$(-\frac{\hbar^2 \nabla^2}{2m} + V_{int} + V_{con})\phi_i = \varepsilon_i \phi_i \tag{1}$$

Herein, $V_{int} + V_{con}$ is the total potential that has the following form :

$$\begin{aligned} V_{int} + V_{con} &= \sum_{LM} V_{LM}(r) Y_{LM}(\hat{r}) \quad \text{muffin-tin sphere} \\ V_{int} + V_{con} &= \sum_{K} V_K e^{i K \cdot r} \quad (V_{con}=0) \quad \text{outside sphere} \end{aligned} \tag{2}$$

where $V_{int}$ the lattice periodic potential that is the sum of the Coulomb and the exchange-correlation potentials.

Fig. 6 draws the optimized structures of monolayer SnSe when different CPs are applied to muffin-tin spheres at fixed lattice constants. With increasing negative/positive CP, the distance between Sn atoms remains 4.38 Å as the same as that between Se atoms. For negative CP case, the distance between Sn and Se atoms becomes short with increasing CP. However, the positive CP can enlarge the distance. When the CP is 0.9 Ry the distance increases from 2.747 Å to 3.204 Å. On the contrary, the distance shorten to 2.468 Å at CP = -0.9 Ry.

It can be seen from Fig. 7 that both PBE and mBJ band gaps decrease with the increase of positive CP, and increase with the increase of negative CP. The PBE band gaps are smaller than the mBJ. At CP = 0 Ry, the mBJ+SOC band gap is 1.0 eV, which is larger than 0.86 eV of SnSe bulk but slightly smaller than the mBJ band gap 1.06 eV. The difference between the band gaps with and without SOC suggests the SOC splitting of electronic structure occur. The difference seems to be larger when the negative CP increases. At CP of above 0.6 Ry, the band gaps are closed and thus monolayer SnSe should show metallic behavior. However, the negative CP enlarges the band gaps, and the mBJ band gaps with and without SOC reach up to 2.8 and 3.2 eV at CP of -1.2 Ry, respectively.

The mBJ band structure as a function of CP is plotted in Fig. 8. At CP of 0 Ry, there are a few nearly linear dispersions along Γ-Y and Γ-X directions. Each two of the linear dispersions touch each other on points D1, D2, D3 and D4, respectively (see Fig. 8(a)). In fact, the nearly linear dispersions have negligible contribution to electronic transport property due to being far away from the band edge. In order to have a conducive contribution to electronic transport, the linear dispersion should meet two conditions: one is the linear dispersion point appears close to band edge (especially VBM and CBM) with a suitable band gap; On the other hand, the dispersion should be linear along all directions. Very few materials fit such harsh conditions. For example, although famous 2D material graphene has the linear dispersion conductivity and valley bands, but it possesses zero band gap, which restricts its wide applications in microelectronics and optoelectronics. Thus, many efforts have been devoted to open the band gap of graphene, but the effect is not fully satisfactory. We find that the touching points (D1, D2, D3 and D4) split and nearly linear dispersions vanish when the negative CP is applied. However, when applying positive CP, the linear dispersions become more obvious and touching points remain. More interestingly, other nearly linear dispersions gradually form. At CP of 0.6 Ry, there are two pairs of nearly linear dispersion respectively located in the $\Gamma$-X and $\Gamma$-Y directions, and both VBM and CBM are linear dispersion peak points. The band gap of 0.21 eV is appears between the pair in the $\Gamma$-Y direction, which is as small as that of 0.20 eV in the $\Gamma$-X direction. At CP of 0.9 Ry, the two band gaps become 0.07 and 0 eV. In addition, two other new pair of linear dispersion in the $\Gamma$-X and $\Gamma$-Y directions also appear.

Theoretically, it is found the SOC usually induces the small opening of band gap and even destroy the linear dispersion of some Dirac material. Thus, we take the impact of SOC on electronic structure into account (see Fig. 9). It is expected that each of the touching points D1, D2, D3 and D4 vanishes, where the linear dispersions become unconspicuous and vanish as the increase of negative CP. However, the linear dispersions induced by the positive CP seem not to be destroyed, although there are large SOC splittings around the linear dispersions (see right column of Fig. 9). At CP of 0 Ry, there are obvious SOC splittings ( ~100 meV) at local VBM and VBM in the $\Gamma$-Y direction but the $\Gamma$-X direction. Interestingly, the SOC splitting fades away when

CP raises up to 0.6 Ry.

That linear dispersion exists in all directions is very import to transport property. In order to verify this, we presented 3D structures of VB and CB in Fig 10. Obviously, the CP could induce large change of band structure shape, and there is a certain degree of electron-hole symmetry especially in band edge although not high as graphene. The electron-hole symmetry becomes high as the positive CP increases. Thus, we only discuss and analyze 3D structures of VB. At CP of 0.6 and 0.9 Ry, notwithstanding no standard Dirac forms, the local VBM locates at the intersection (the white and red in Fig. 10(e)) of two quasi-planes. This implies the two quasi-planes except for the intersection are nearly massless. These provide possibility to manipulate the nature of electrical transport. One can apply different CP to tune the band gap and the shape of band edge, and find suitable balance between the band gap and linear dispersion.

## IV. SUMMARY

In summary, we use the first-principles method to study the effects of biaxial strain and CP on the electronic structure of monolayer SnSe. It is found that compressive strain can narrow band gap due to the introduction of high dispersion of valley bands, and tensile strain reduces dispersion of valley bands and thus enlarge the band gap. Also, the band gaps can be tuned by applying CP. At CP of 0.9 Ry, there are nearly linear dispersions existing in VB and CB, which are not destroyed by SOC. Our study is in favor of understanding the mechanism of the Dirac cones and proposes that applying CP is new design strategy for electronic structure of 2D materials. Although this method may be difficult to implement experimentally, it is a worthwhile attempt.

**ACKNOWLEDGMENT:**
This work is supported by National Natural Science Foundation of China (Grant No. 11804132), National Natural Science Foundation of China (Grant No. 11847129) and Sichuan Science and Technology Program (Grant No. 2019YJ0336).

**Author contributions**

A. J. H. and C. L. Y. conceived this research. A. J. H. performed all the computations and wrote the manuscript. All the authors discussed the results and commented on the paper.

[1] K. S. Novoselov, A. K. Geim, S. V. Morozov, D. Jiang, Y. Zhang, S. V. Dubonos, I. V. Grigorieva, and A. A. Firsov, Science **306**, 666 (2004).
[2] X. D. Xu, W. Yao, D. Xiao, and T. F. Heinz, Nature Physics **10**, 343 (2014).
[3] P. Vogt, P. De Padova, C. Quaresima, J. Avila, E. Frantzeskakis, M. C. Asensio, A. Resta, B. Ealet, and G. Le Lay, Phys. Rev. Lett. **108**, 155501 (2012).
[4] F. F. Zhu, W. J. Chen, Y. Xu, C. L. Gao, D. D. Guan, C. H. Liu, D. Qian, S. C. Zhang, and J. F. Jia, Nat. Mater. **14**, 1020 (2015).
[5] S. Z. Butler, S. M. Hollen, L. Y. Cao, Y. Cui, J. A. Gupta, H. R. Gutierrez, T. F. Heinz, S. S. Hong, J. X. Huang, A. F. Ismach, E. Johnston-Halperin, M. Kuno, V. V. Plashnitsa, R. D. Robinson, R. S. Ruoff, S. Salahuddin, J. Shan, L. Shi, M. G. Spencer, M. Terrones, W. Windl, and J. E. Goldberger, Acs Nano **7**, 2898 (2013).
[6] S. P. Koenig, R. A. Doganov, H. Schmidt, A. H. C. Neto, and B. Ozyilmaz, Appl. Phys. Lett. **104**, 103106 (2014).
[7] J. S. Qiao, X. H. Kong, Z. X. Hu, F. Yang, and W. Ji, Nat. Commu. **5**, 4475 (2014).
[8] H. B. Ribeiro, M. A. Pimenta, C. J. S. de Matos, R. L. Moreira, A. S. Rodin, J. D. Zapata, E. A. T. de Souza, and A. H. C. Neto, Acs Nano **9**, 4270 (2015).
[9] L. C. Gomes and A. Carvalho, Phys. Rev. B **92**, 085406 (2015).
[10] J. D. Wiley, A. Breitschwerdt, and E. Schönherr, Solid State Commun. **17**, 355 (1975).
[11] R. Eymard and A. Otto, Phys. Rev. B **16**, 1616 (1977).
[12] D. Z. Tan, H. E. Lim, F. J. Wang, N. B. Mohamed, S. Mouri, W. J. Zhang, Y. Miyauchi, M. Ohfuchi, and K. Matsuda, Nano Research **10**, 546 (2017).
[13] D. D. Vaughn, R. J. Patel, M. A. Hickner, and R. E. Schaak, J. Am. Chem. Soc. **132**, 15170 (2010).
[14] A. J. Hong, L. Li, H. X. Zhu, Z. B. Yan, J. M. Liu, and Z. F. Ren, J. Mater. Chem. A **3**, 13365 (2015).
[15] L. D. Zhao, S. H. Lo, Y. S. Zhang, H. Sun, G. J. Tan, C. Uher, C. Wolverton, V. P. Dravid, and M. G. Kanatzidis, Nature **508**, 373 (2014).
[16] A. Ramasubramaniam, D. Naveh, and E. Towe, Phys. Rev. B **84**, 205325 (2011).
[17] A. A. Kistanov, S. K. Khadiullin, S. V. Dmitriev, and E. A. Korznikova, Chem. Phys. Lett. **728**, 53 (2019).
[18] Y. Ishikawa, K. Wada, D. D. Cannon, J. F. Liu, H. C. Luan, and L. C. Kimerling, Appl. Phys. Lett. **82**, 2044 (2003).
[19] Z. H. Dai, L. Q. Liu, and Z. Zhang, Adv. Mater. **31**, e1805417 (2019).
[20] L. N. Du, C. Wang, W. Q. Xiong, S. Zhang, C. X. Xia, Z. M. Wei, J. B. Li, S. Tongay, F. Y. Yang, X. Z. Zhang, X. F. Liu, and Q. Liu, 2d Materials **6**, 025014 (2019).
[21] D. M. Hoat, T. V. Vu, M. M. Obeid, and H. R. Jappor, Chem. Phys. **527** (2019).
[22] J. Y. Shi, Y. Ou, M. A. Mighorato, H. Y. Wang, H. Li, Y. Zhang, Y. S. Gu, and M. Q. Zou, Comp. Mater. Sci. **160**, 301 (2019).
[23] C. H. Zhang, G. Xiang, M. Lan, and X. Zhang, Chinese Physics B **23**, 096103 (2014).
[24] Y. B. Zhang, T. T. Tang, C. Girit, Z. Hao, M. C. Martin, A. Zettl, M. F. Crommie, Y. R. Shen, and F. Wang, Nature **459**, 820 (2009).
[25] W. Chen, Y. F. Li, G. T. Yu, C. Z. Li, S. B. B. Zhang, Z. Zhou, and Z. F. Chen, J. Am. Chem. Soc. **132**, 1699 (2010).

[26] P. Chen, X. Zhao, T. X. Wang, X. Q. Dai, and C. X. Xia, Solid State Commun. **230**, 35 (2016).
[27] P. Lu, X. J. Wu, W. L. Guo, and X. C. Zeng, Phys. Chem. Chem. Phys. **14**, 13035 (2012).
[28] K. Schwarz, P. Blaha, and G. K. H. Madsen, Comput. Phys. Commun. **147**, 71 (2002).
[29] J. P. Perdew, K. Burke, and M. Ernzerhof, Phys. Rev. Lett. **77**, 3865 (1996).
[30] F. Tran and P. Blaha, Phys. Rev. Lett. **102**, 226401 (2009).
[31] E. Sjöstedt, L. Nordström, and D. J. Singh, Solid State Commun. **114**, 15 (2000).

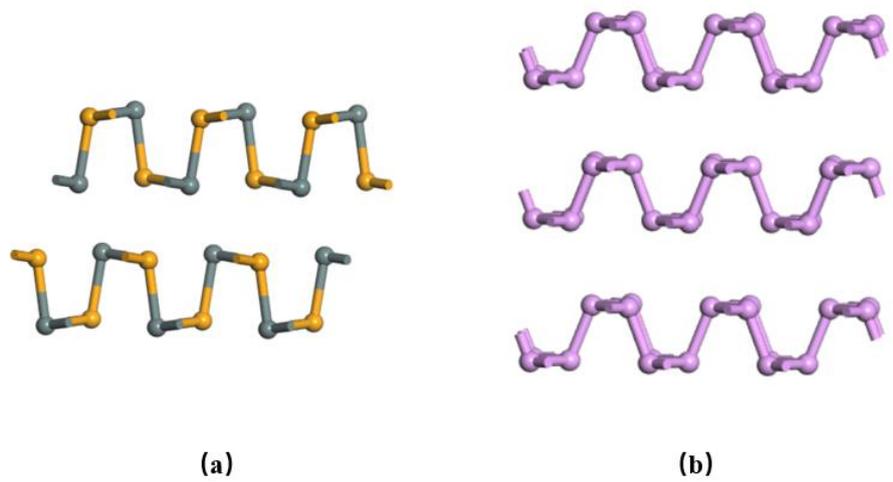

Fig. 1 (Color online) Atomic structures of low-temperature phase SnSe (a) and black phosphorus (b).

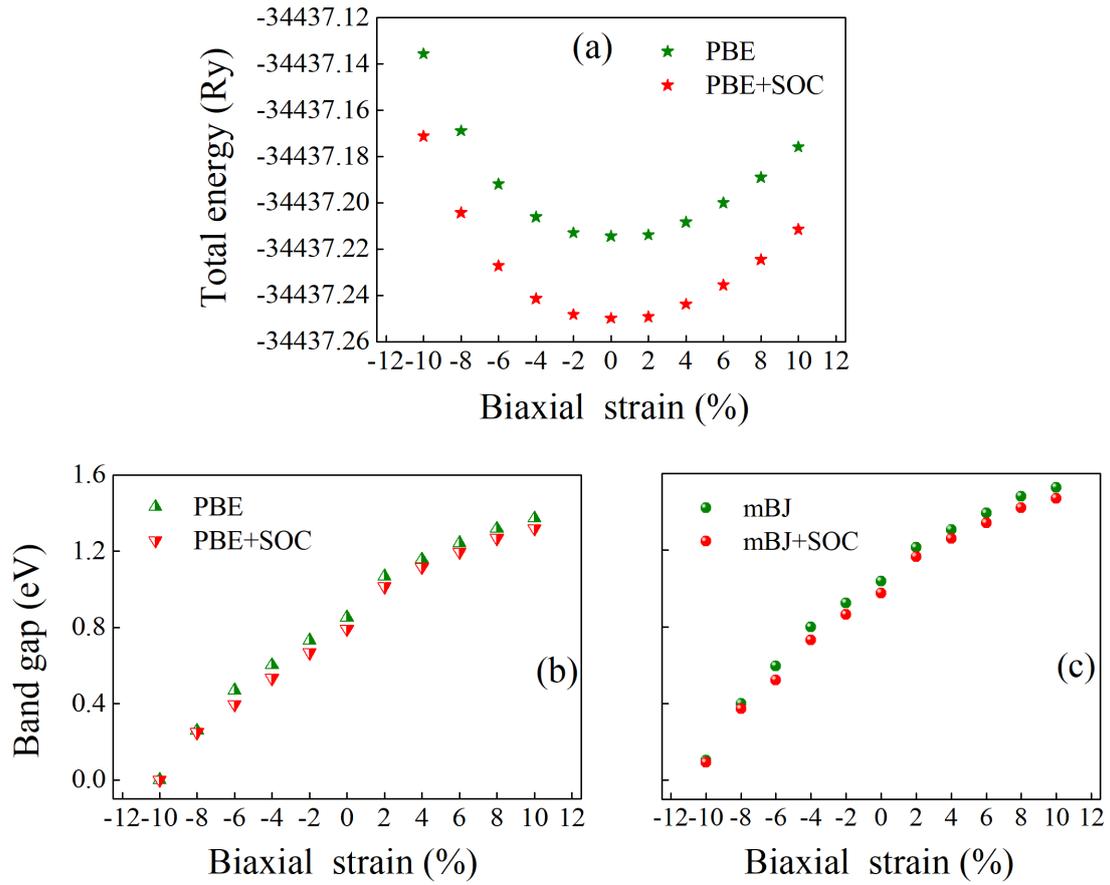

Fig. 2 (Color online) Total energy as a function of biaxial strain (a), PBE (b) and mBJ (c) band gaps with and without SOC as a function of biaxial strain.

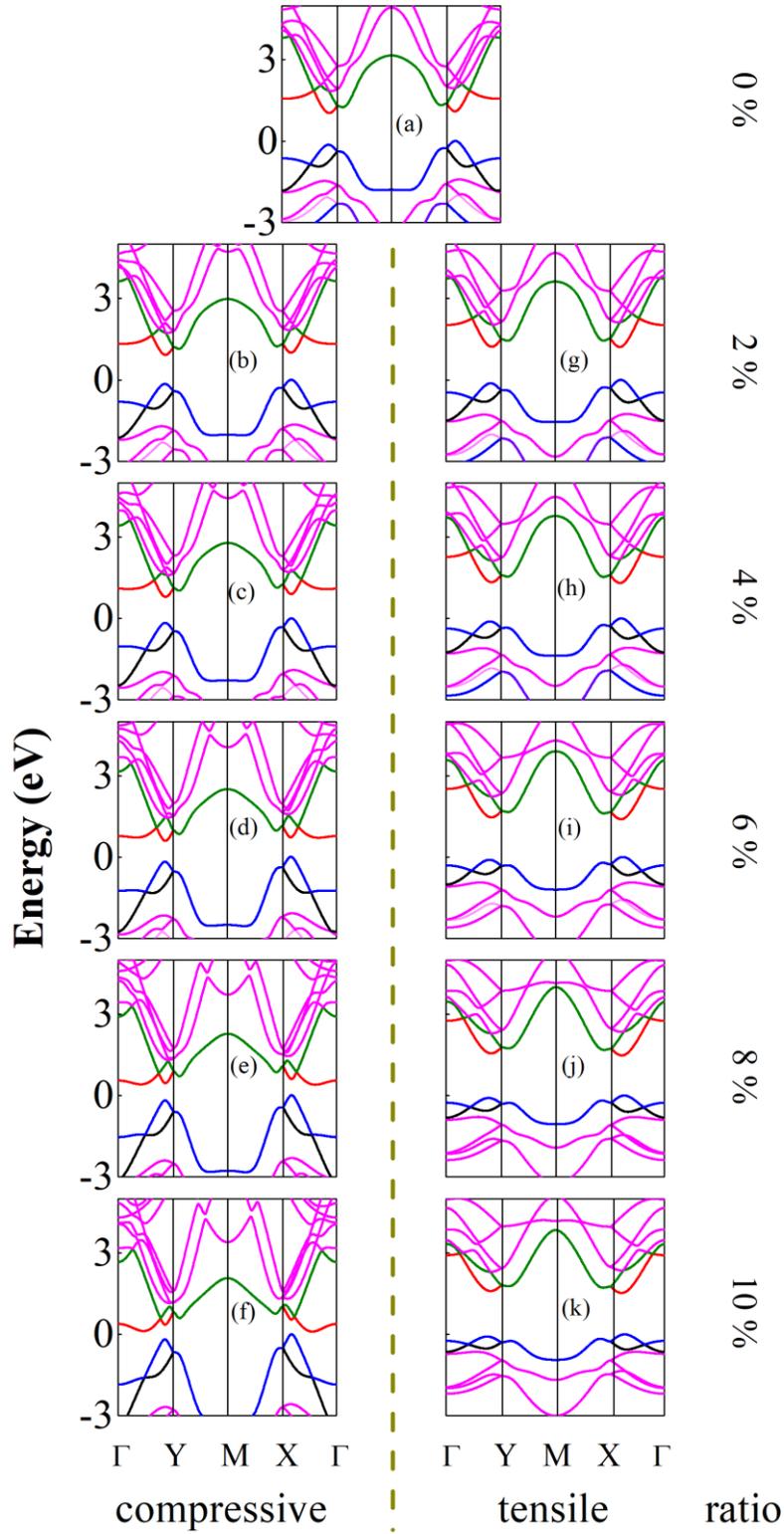

Fig. 3 (Color online) mBJ band structure without SOC as a function of biaxial strain in monolyer SnSe.

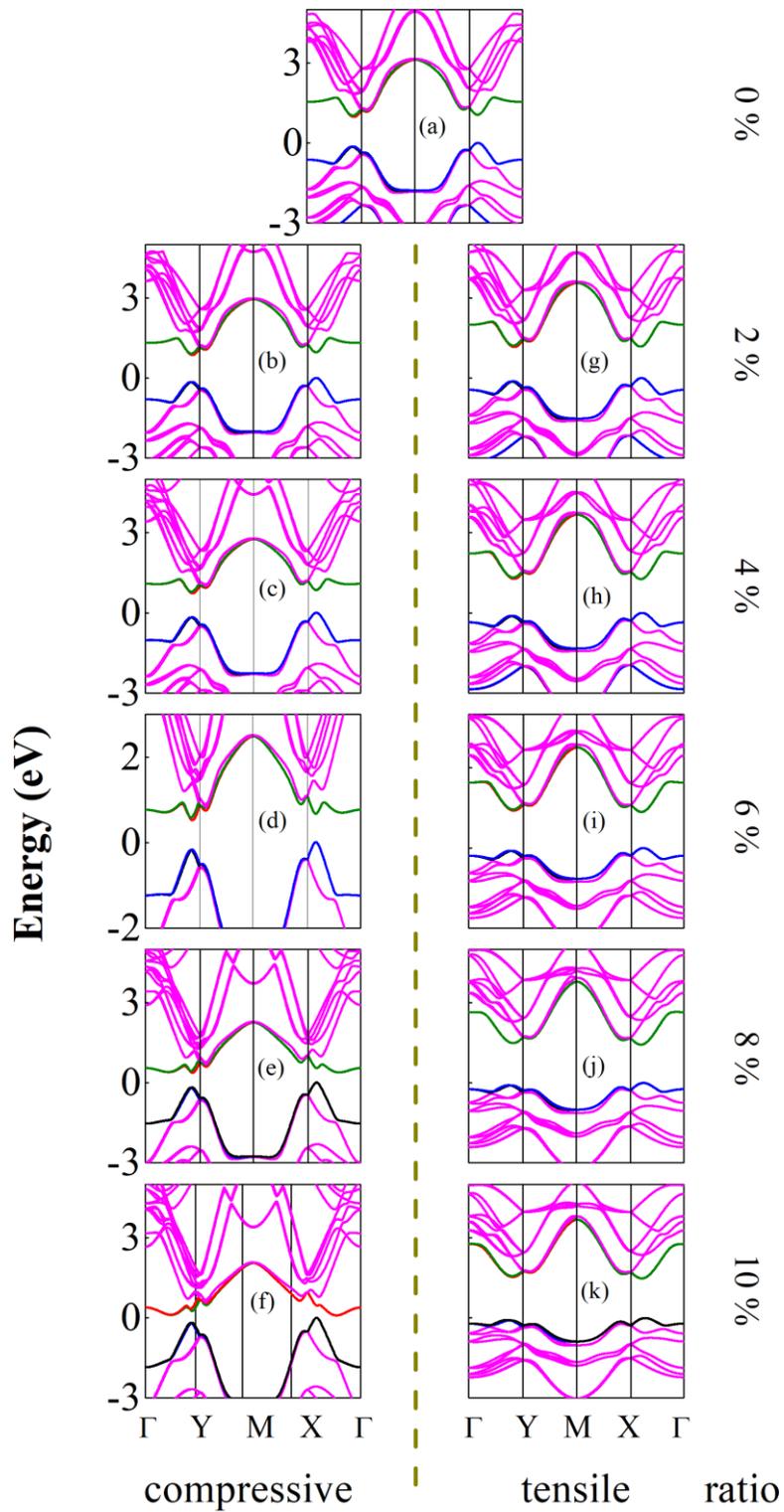

Fig. 4 (Color online) mBJ band structure with SOC as a function of biaxial strain in monolyer SnSe.

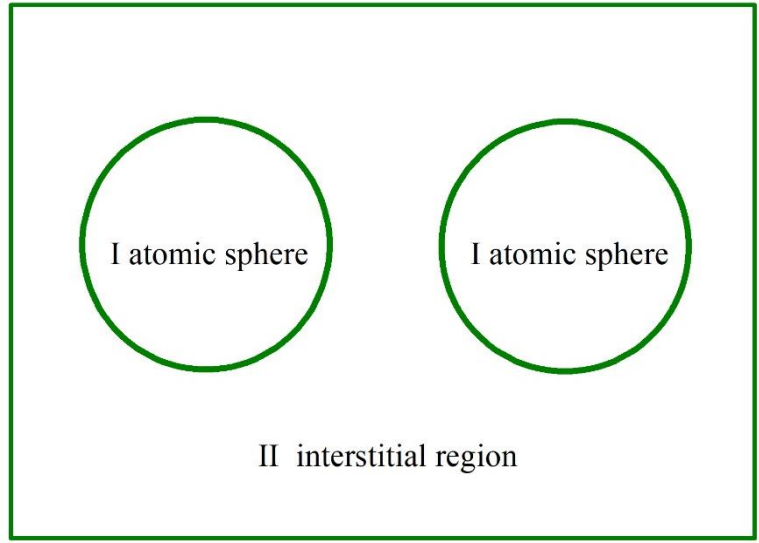

Fig. 5 (Color online) Sketch of partitioning the unit cell into atomic spheres (I) and an interstitial region (II).

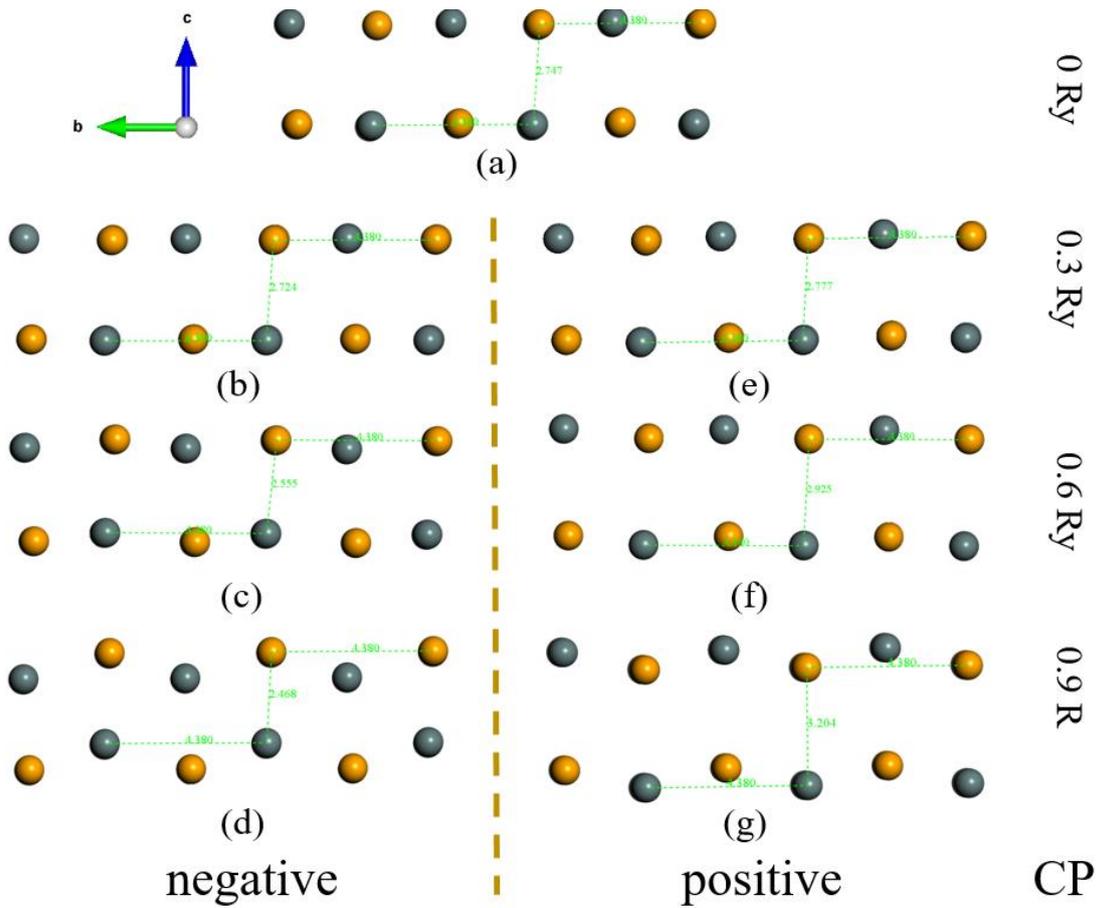

Fig. 6 (Color online) Optimized atomic structures of monolayer SnSe with applying different CPs.

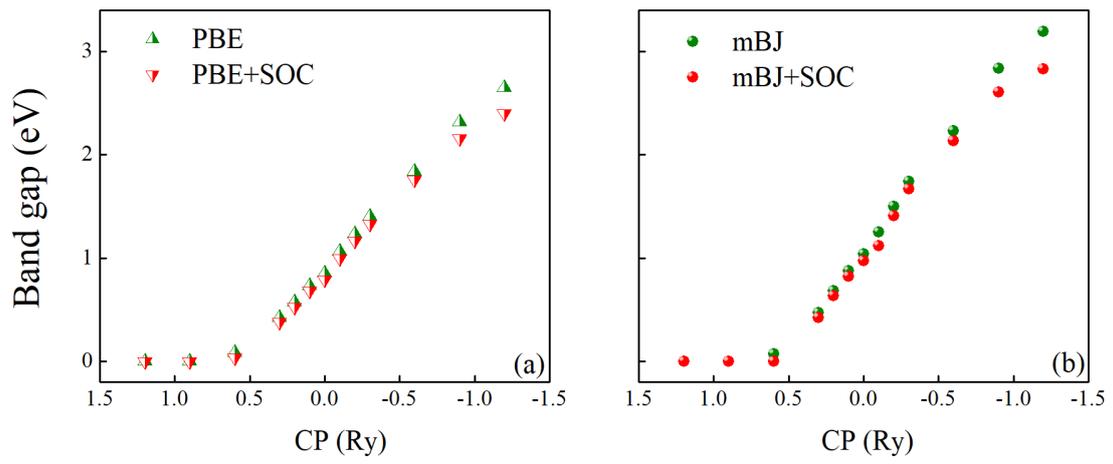

Fig. 7 (Color online) PBE (a) and mBJ (b) band gaps with and without SOC as a function of constant potential.

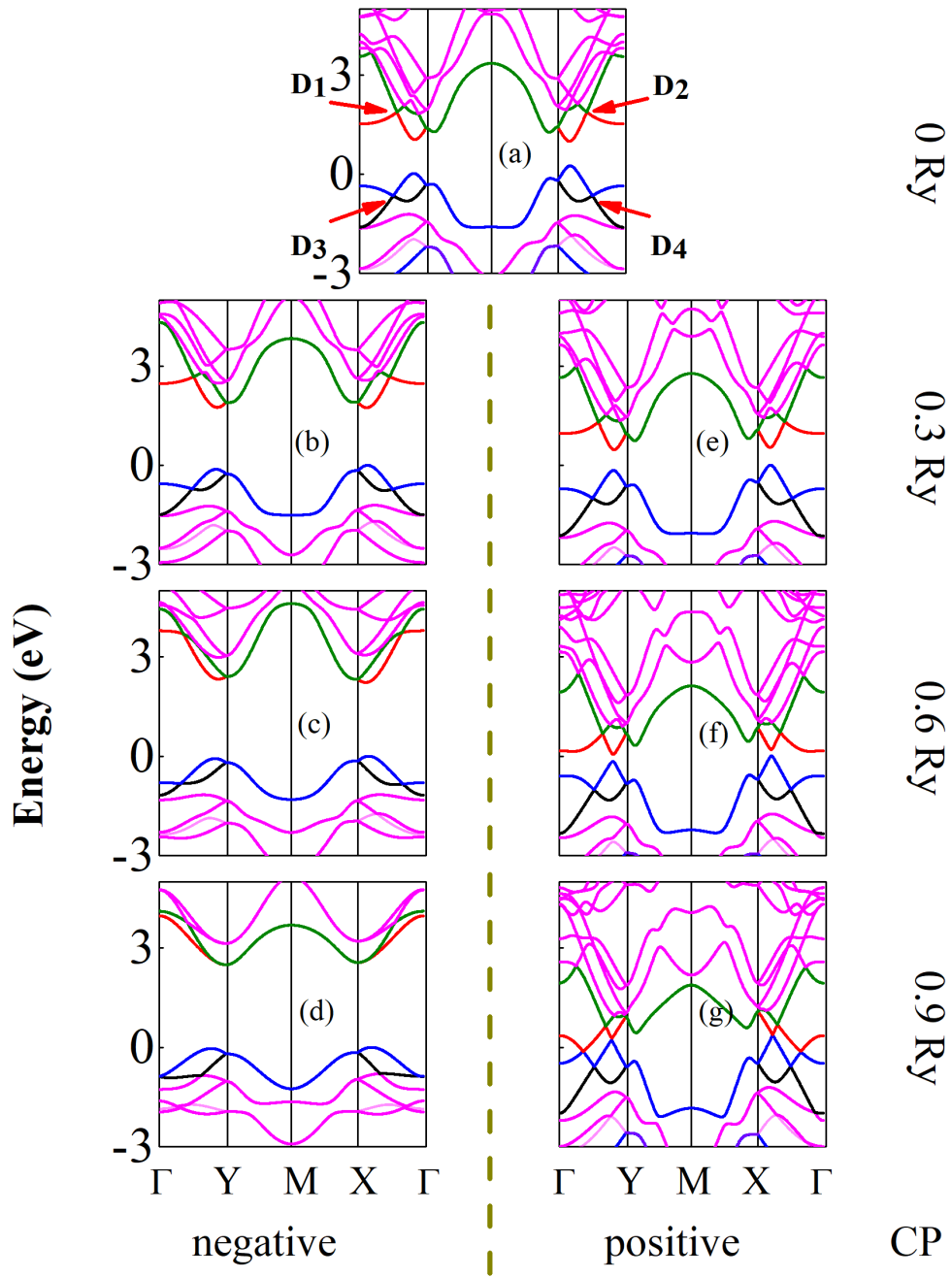

Fig. 8 (Color online) mBJ band structure with SOC as a function of CP in monolyer SnSe.

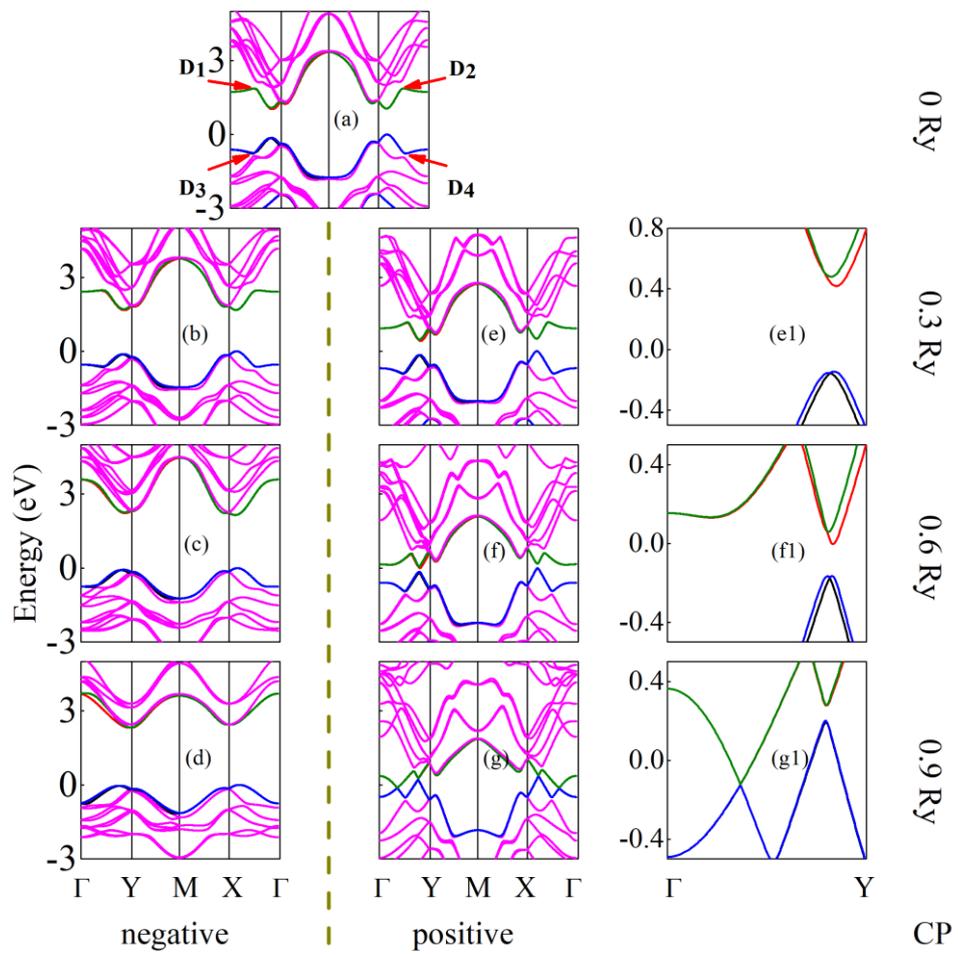

Fig. 9 (Color online) mBJ band structure with SOC as a function of CP in monolyer SnSe. Right column: enlarged band structures along the Γ-Ydirection at positive CP.

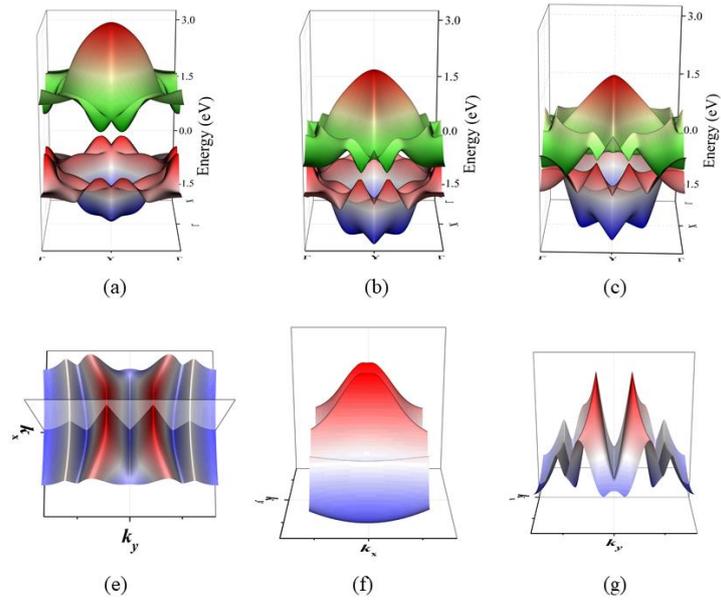

Fig. 10 (Color online) 3D band structures of mBJ+SOC valence and conductivity bands at different external constant potentials of (a) 0 (b) 0.6 and (c) 0.9 Ry. (e)-(g) Top and side views of 3D structures of valence band near VBM.